\documentclass[reprint,amsmath,amssymb,aps,pra,twocolumn,10pt,groupedaddress,floatfix,noeprint,nofootinbib]{revtex4-2}
\usepackage{babel}
\usepackage{tabularray}
\usepackage{graphicx}
\usepackage{dcolumn}
\usepackage{bm}
\usepackage{physics}
\usepackage[caption=true]{subfig}
\usepackage{overpic}
\usepackage{hyperref} 
\usepackage{soul}

\hypersetup
{
colorlinks=true,       
linkcolor=blue,          
citecolor=blue,        
filecolor=blue,      
urlcolor=blue           
}

\usepackage{xcolor}

\newcommand{\avg}[1]{\langle #1 \rangle}

\newcommand{\mr}[1]{\mathrm{#1}}
\newcommand{\aop}{\ensuremath{\hat{a}}}
\newcommand{\adop}{\ensuremath{\hat{a}^{\dagger}}}

\newcommand{\Hop}{\hat{H}}

\newcommand{\Jop}{\hat{J}}

\newcommand{\hrho}{\hat{\rho}}

\newcommand{\pvedit}[1]{\textcolor{black}{#1}}

\begin{document}

\title{Comparative Study of Indicators of Chaos in the Closed and Open Dicke Model}

\author{Prasad Pawar}
\email{prasad.pawar@iitgn.ac.in}
\author{Arpan Bhattacharyya}%
\email{abhattacharyya@iitgn.ac.in}
\author{B. Prasanna Venkatesh}
\email{prasanna.b@iitgn.ac.in}
\affiliation{Indian Institute of Technology Gandhinagar, Palaj, Gujarat 382355, India}

\begin{abstract} 
The Dicke model, renowned for its superradiant quantum phase transition, also exhibits a transition from regular to chaotic dynamics. In this work, we provide a systematic, comparative study of static and dynamical indicators of chaos for the closed and open Dicke model. In the closed Dicke model, we find that indicators of chaos sensitive to long-range correlations in the energy spectrum, such as the spectral form factor (SFF), can deviate from the Poissonian predictions and show a \emph{dip-ramp-plateau} feature even in the regular region of the Dicke model unless very large values of the spin size are chosen. Thus, care is needed in using such indicators of chaos in general. In the open Dicke model with cavity damping, we find that the dissipative spectral form factor emerges as a robust diagnostic displaying a quadratic dip-ramp-plateau behavior in agreement with the Ginibre Unitary Ensemble (GinUE) in the superradiant regime. Moreover, by examining the spectral properties of the Liouvillian, we provide indirect evidence for the concurrence of the dissipative superradiant quantum phase transition and the change in Liouvillian eigenvalue statistics from 2-D Poissonian to GinUE behavior.
\end{abstract}

\maketitle

\section{Introduction}\label{sec:Intro}

The study of quantum chaos has garnered significant attention, particularly in relation to the spectral and dynamical features of quantum systems exhibiting chaotic behavior. A key characteristic of chaotic quantum systems is the statistical distribution of their energy levels, which can often be described using Random Matrix Theory (RMT). Originally formulated by Wigner to explain the spectral properties of complex many-body systems such as atomic nuclei \cite{Wigner_1967,Guhr_1998}, RMT predicts universal statistical properties that distinguish chaotic quantum systems from integrable ones. Bohigas, Giannoni, and Schmit \cite{Bohigas_1984} first proposed that the spectral properties of quantum systems whose classical counterparts exhibit chaos align with those of random matrix ensembles, a result that is known in the community as the Bohigas-Giannoni-Schmit (BGS) conjecture. In particular, the nearest-neighbor spacing distribution (NNSD) \cite{Bohigas_1984, Guhr_1998, Emary_2003, Gomez_2002, Abul_Magd_2014}, which shows Poissonian (uncorrelated spectrum) like behavior for (quasi-)integrable systems reflecting clustering of energy levels as opposed to chaotic systems that adopt a Wigner-Dyson distribution \cite{Guhr_1998}, has emerged as a popular and robust indicator of quantum chaos. More recently, the NNSD has also been successfully extended to the study of chaos in open quantum systems described by Lindblad master equations where the role of the energy eigenvalues is replaced by the complex eigenvalues of the Liouvillian \cite{PhysRevLett.123.090603, Hamazaki_2020, sa_complex_2020, hamazaki_2022, Grobe_1988, Prasad_2022}. Apart from the NNSD, a closely related indicator is the level spacing ratio of either the nearest neighbor levels or those separated by $k$ other energy levels (the so called k$^\mr{th}$ level spacing ratio), which is particularly useful as it does not require the unfolding procedure needed for NNSD analysis \cite{Oganesyan_2007,Atas_2013,kota2014embedded,Tekur_2018, 2020PhRvB.102e4202R, sa_complex_2020}.

While the NNSD and other level spacing ratios are static measures of quantum chaos, there has been an effort to also identify dynamic or time-dependent indicators of quantum chaos \cite{otoc, Leviandier_1986,Shenker:2013pqa,Maldacena:2015waa,otoc1,Chowdhury:2017jzb}. Chief among such measures is the spectral form factor (SFF) \cite{Leviandier_1986,PhysRevLett.67.1185,PhysRevA.46.4650,PhysRevLett.70.572,Brezin_1997, Bao_2015, Cotler_2017,SFFKickedTop,SFFBilliards} that is defined as the Fourier transform of the two-point energy correlation function. While NNSD and nearest neighbor level spacing ratios capture short-range correlations, the SFF is sensitive to both short and long-range correlations, making it a more comprehensive diagnostic tool \cite{Cotler_2017} for level repulsion in the spectrum. An attractive property of the SFF as an indicator of chaos arises from the fact that random matrix models such as the Gaussian Orthogonal Ensemble (GOE) exhibit a characteristic \textit{dip-ramp-plateau} structure in the SFF, sometimes also known as the correlation hole, allowing one to identify the presence of quantum chaos in general Hamiltonians by comparison  \cite{Brezin_1997, Papadodimas:2015xma,Garcia-Garcia:2016mno, Krishnan:2016bvg,Dyer:2016pou, Cotler_2017, delCampo:2017bzr, PhysRevD.100.026017,Winer:2022ciz, Winer_2023, Okuyama:2023pio,Cipolloni_2023,Matsoukas-Roubeas_2023,Bhattacharyya_2023_SFF,Bhattacharyya:2023gvg,Das_2024}. Moreover, by viewing the SFF as 
the survival probability of the Coherent Gibbs State (CGS) under Hamiltonian time evolution \cite{Torres-Herrera_2018, Villaseñor_2020, Lerma-Hernandez_2019}, allows a direct extension to (Markovian) open quantum systems in the form of the dissipative survival probability function (DSPF) where the time-evolution is now generated by the associated Liouvillian operator \cite{Tameshtit_1992, Torres-Herrera_2018, delCampo_2017, Xu_2021, Apollonas_2023, Cornelius_2022, Matsoukas-Roubeas_2023}. In addition, a dissipative spectral form factor (DSFF) can also be defined for open quantum systems by considering the eigenvalues of the Liouvillian and constructing a function analogous to the SFF by taking a 2D Fourier transform of the eigenvalue correlations in the complex plane \cite{Li_2021,Li_2024}.

Since a universal measure or indicator of quantum chaos is still elusive \cite{MichaelBerry_1989, hunterjones2018chaos, Aurich_1994}, it is important to find specific quantum systems to benchmark and comparatively study different measures of chaos. In this context, the Dicke model, which describes the interaction between a single-mode bosonic field and a large ensemble of two-level atoms, has long served as a very useful testbed for studying quantum chaos \cite{Bastarrachea_Magnani_2016, Emary_2003,Emary_2003_prl,Bao_2015,Villasenor_2023, Bhattacharya_2015, Tiwari_2023, Wang_2022, Wang_2020, Pilatowsky-Cameo2021, Lerma-Hernandez_2019, Villasenor_2024_open1, Villasenor_2024_open2, Prasad_2022, Larson_2017,Prasad_2024,Vivek_2025}. The Dicke model undergoes a quantum phase transition (QPT) from a normal-to-superradiant phase when the atom-field coupling exceeds a critical value, and more interestingly, this transition was also shown to be concurrent with a transition of the NNSD from Poisson to Wigner-Dyson behavior \cite{Emary_2003}. Further careful studies have also shown that the superradiant phase has spectral properties comparable to those of the GOE. Given that the classical limit of the Dicke model in the superradiant regime is chaotic, this serves as a validation of the BGS conjecture \cite{Emary_2003, Wang_2020, Wang_2022, Bao_2015}. While these works primarily focus on NNSD and level spacing ratios as indicators of chaos, the SFF, especially over the entire range of coupling strengths of the Dicke model, remains relatively underexplored. In this context, focusing on the related measure of survival probability, the development of the correlation hole and its comparison to GOE in the superradiant regime was shown in \cite{Lerma-Hernandez_2019} and 
a comprehensive study across normal and superradiant regimes especially the dependence on the initial state and comparison to classical dynamics was given in \cite{Villaseñor_2020}. On the dissipative front, it is known that the Dicke model with cavity decay also has a dissipative quantum phase transition \cite{Dimer_2007} from normal to superradiant regimes. Studies examining quantum chaos within this model \cite{Prasad_2022} have shown that the NNSD of the Liouvillian eigenvalues changes from 2D Poissonian  in the normal regime to that of the Ginibre Unitary Ensemble (GinUE) deep in the superradiant regime. Additionally, recent work \cite{Villasenor_2024_open1, Villasenor_2024_open2} suggests that while the complex spacing ratio of the open Dicke model exhibits GinUE behavior in the superradiant regime, its classical counterpart is not chaotic in the steady state. Furthermore, the relation between GinUE behavior and classical chaos is better understood via the concept of transient chaos introduced in \cite{Mondal_2025}. These works provide a timely reminder that validating extensions of the BGS conjecture to open quantum systems (i.e., Grobe-Haake-Sommers (GHS) conjecture by \cite{Grobe_1988}) may not be very straightforward.

In this work, we provide a systematic comparative study of static and dynamical indicators of chaos for the closed and open Dicke model. In particular, we have studied the (see Table \ref{tab:acronyms}) NNSD, k$^{\mr{th}}$-level spacing ratio and the SFF in the closed Dicke model and the NNSD, complex level spacing ratio, and DSFF in the open case. Our central findings are as follows. In the closed Dicke model, while the NNSD and nearest neighbor level spacing ratio show a clear transition from Poissonian behavior to that of GOE RMT behavior, as expected from previous results, the SFF and k$^{\mr{th}}$-level spacing ratio require more careful consideration. Specifically, we find that the k$^{\mr{th}}$-level spacing ratio can deviate from the Poissonian predictions, and SFF can show a dip-ramp-plateau-like behavior even in the regular region of the Dicke model for any finite values of the spin size $j=N/2$. Thus, the long-range energy correlations responsible for this observed behavior in the regular region persist unless we take the ultimate thermodynamic limit $N \rightarrow \infty$. In the open Dicke model, by examining the spectral properties of the Liouvillian for different values of the cavity damping, we provide strong indirect evidence for the concurrence of the dissipative superradiant quantum phase transition and the change in NNSD from 2-D Poissonian to GinUE RMT behavior. Finally, we calculate the DSFF for the open Dicke model using the unfolding of complex spectra introduced in \cite{Li_2024} and show clearly that the quadratic dip-ramp-plateau behavior in agreement with the GinUE RMT emerges in the superradiant regime.

The paper is organized as follows. We introduce the Dicke model in both closed and open settings and describe the relevant RMT ensembles, GOE and GinUE in Sec.~(\ref{sec:DickeModel&RMT}). In Sec.~(\ref{sec:chaos measures}) we provide an overview of different indicators of chaos that will be analysed for both open and closed models. Finally, we present our results and conclusions in Sec.~(\ref{sec:results}) and Sec.~(\ref{sec:conclusion}) respectively. Some additional details and results not covered in the main paper are presented in the appendix~(\ref{app:A}), (\ref{app:B}) and (\ref{app:C}).

\begin{table}
    \centering
   \begin{tblr}{|p{1.5cm}|p{6cm}|}
        \hline
        \textbf{Acronym} & \hspace{2cm}\textbf{Full Form}\\
        \hline
        CGS & Coherent Gibbs State (see Eq.~\eqref{eq:CGSState})\\
        \hline
        SFF & Spectral Form Factor (see Eq.~\eqref{eq:SFF_defn_1})\\
        \hline
        DSFF & Dissipative Spectral Form Factor (see Eq.~\eqref{eq:dsff_def1})\\
        \hline
        DSPF & Dissipative Survival Probability Function (see Eq.~\eqref{eq:DSPF})\\
        \hline
        GinUE & Ginibre Unitary Ensemble (matrices with elements $\in N_\mathbb{C}(0,1)$)\\
        \hline
        GOE & Gaussian Orthogonal Ensemble (symmetric matrices with real elements $\in N(0,1)$)\\
        \hline
        NNSD & Nearest-Neighbor Spacing Distribution \\
        \hline
    \end{tblr}
    \caption{Quick reference table for important acronyms frequently used in the text.}
    \label{tab:acronyms}
\end{table}

\section{Dicke Model}
\label{sec:DickeModel&RMT} 
The Dicke model  describes a light-matter system of $N$ spin-1/2 particles interacting collectively with the single bosonic electromagnetic field mode of an ideal cavity with the Hamiltonian \cite{PhysRev.93.99} (we take $\hbar = 1$ throughout),
\begin{align}
\Hop = \omega_0 \Jop_z + \omega \aop^\dagger \aop + \frac{g}{\sqrt{2j}}(\aop+\adop)(\Jop_+ + \Jop_-) \label{eq:DickeH},
\end{align}
with $\Jop_z$ denoting the $z-$component of the collective atomic spin of length $j=N/2$, $\Jop_{\pm}$ the corresponding collective ladder operators, and $\aop$ ($\adop$) the annihilation (creation) operator of the cavity mode. The Dicke model hosts a discrete $\mathbb{Z}_2$ parity symmetry with the parity operator \cite{Emary_2003} $\hat{\Pi}=\exp\left[i\pi(\hat{J_z}+\hat{a}^\dagger \hat{a}+j)\right]$. In the thermodynamics limit, $j\rightarrow\infty$, when the coupling $g$ is tuned above the critical value $g_c=\sqrt{{\omega\omega_0}/{2}}$, this parity symmetry is broken, leading to the famous Dicke phase transition from the normal phase to the superradiant phase. The order parameters, given by the atomic ($\avg{\Jop_++\Jop_-}$) and photonic coherence ($\avg{a+\adop}$) as well as the cavity photon number ($\avg{\hat a^\dagger \hat a}$) take non-zero values for the ground state in the superradiant phase and zero values in the normal phase. More interestingly, it was shown in \cite{Emary_2003}, that this normal-to-superradiant phase transition is accompanied by a regular-to-chaotic phase transition of the model. Specifically, both the classical limit of the Dicke model and the spectral statistics characterized by the nearest neighbor spacing distribution (NNSD), to be discussed in detail in Sec.~(\ref{sec:chaos measures}), show a transition from regular to chaotic behavior as $g$ exceeds $g_c$.

In addition to the closed Dicke model, where the energy spectrum of the Hamiltonian defined in Eq.~\eqref{eq:DickeH} completely determines the different indicators of chaos, we are also interested in the open Dicke model with the cavity mode damped at a rate $\gamma$, leading to photon loss from the cavity. One can model this dissipative scenario by the usual Gorini-Kossakowski-Sudarshan-Lindblad (Lindblad for short) master equation for the density matrix $\hrho$ which is given as
\begin{align}
    \dv{\hrho}{t} &= \mathcal{L}[\rho]  \nonumber\\
    &=-i\comm{\Hop}{\rho} + \gamma\left(2\aop\hrho \adop - \adop \aop \hrho -\hrho \adop \aop\right)
    \label{eq:master-eq-Dicke}
\end{align}
$\mathcal{L}[\cdot]$ denotes the Liouvillian super-operator. Similar to the closed case, there is a normal-to-superradiant dissipative quantum phase transition for the open Dicke model with the critical coupling strength of \cite{Dimer_2007}
\begin{align}
    g_{c\gamma} = \frac{1}{2}\sqrt{\frac{\omega_0}{\omega}\left(\gamma^2+\omega^2\right)}.
    \label{eq:gc_cav_damp}
\end{align}
Note that the critical coupling strength in the open case is larger than in the closed case. As we discussed in the introduction, the study of chaos in the open Dicke model is incipient, with some key results presented in  \cite{Prasad_2022,Villasenor_2024_open2}. We will discuss these results when we compare them to our findings in Sec.~(\ref{sec:results}). 

Finally, as discussed earlier, we will also compare our results with those from random matrix ensembles, which are fundamental constructs in random matrix theory, representing families of matrices whose elements are random variables \cite{Wigner_1967, Guhr_1998}. In particular, we will consider the Gaussian Orthogonal Ensemble (GOE) and the Ginibre Unitary Ensemble (GinUE) \cite{Edelman_2005, Bohigas_1984, Guhr_1998, akemann2024, Shivam_2023, Prasad_2022} which have long served as cornerstone models to study quantum chaos (see Table \ref{tab:acronyms} for a brief definition of GOE and GinUE as well as a list of acronyms used in the manuscript). 

\section{Indicators of Chaos - Brief Recapitulation}\label{sec:chaos measures}

The spectral properties and dynamics of quantum systems offer critical insights into chaotic behavior. In order to make our presentation self-contained and clear, we now provide a brief account of different indicators of chaos used in closed and open quantum systems. We begin with measures directly based on the statistical behavior of the spectrum that relies on the tendency of the energy levels of a chaotic quantum system to repel each other \cite{Cotler_2017,Bohigas_1984} and follow it up with dynamical indicators such as the spectral form factor. 

\begin{figure*}
    \centering{
\subfloat{\begin{overpic}[abs,width=0.475\linewidth]{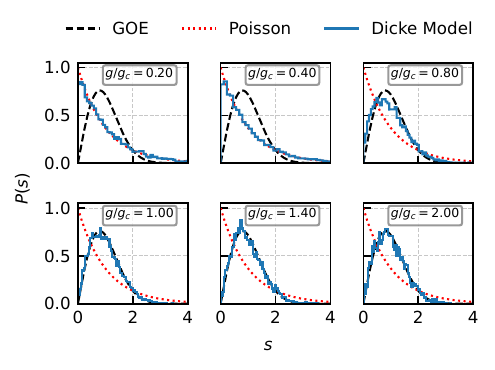}
\put(85,89){\textcolor{black}{\textbf{(a)}}}
\end{overpic} \label{pic1a}
}
\subfloat{\begin{overpic}[abs,width=0.475\linewidth]{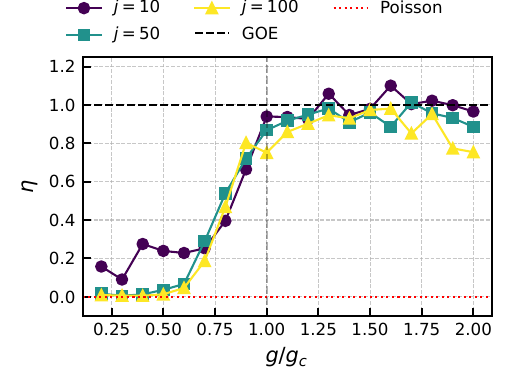}
\put(130,85){\textcolor{black}{\textbf{(b)}}}
\end{overpic}\label{pic1b}
}}
\caption{(a) NNSD for the Dicke Model (closed) as a function of the coupling strength $g/g_c$. The black dashed (red solid) lines depict the NNSD for the GOE (Poisson) RMT. (b) Normalized distance of the Dicke model NNSD from the GOE distribution, $\eta$, as a function of scaled coupling $g/g_c$ and different values of $j$. We have taken $j=50$ in (a) and photon cutoff $M=400$ in (a,b).}
\label{fig:fig1}   
\end{figure*}
\subsection{Spectral Indicators of Chaos}
Considering closed quantum systems first, the most prevalent indicator of chaos \cite{Emary_2003, Wang_2022, Bao_2015} is the nearest-neighbor spacing distribution (NNSD) of energy levels, denoted by $P(s)$. Here, the variable $s$ denotes the spacing between the nearest unfolded energy levels $\mathcal{E}_{i}$ and $\mathcal{E}_{i+1}$ \emph{i.e.} $s_i = \mathcal{E}_{i+1} - \mathcal{E}_{i}$. We use standard procedures to unfold the energy levels of a hamiltonian by first fitting smooth functions, such as polynomials \cite{Guhr_1998,Abul_Magd_2014,Gomez_2002}, to the numerically calculated spectral density.

In the open case, the chaotic nature of the Dicke model is determined by the Liouvillian as mentioned in Eq.~\eqref{eq:master-eq-Dicke}. In \cite{Prasad_2022}, the NNSD for the complex eigenvalue spectrum of the Liouvillian $\mathcal{L}$ for the Dicke model was analyzed. Since we aim to compare this NNSD behavior against other indicators of chaos, we briefly recall the procedure used to compute it. The idea is to first calculate the Euclidean distance between each of the complex eigenvalues $\lambda_i$ of $\mathcal{L}$ and its nearest neighbor $s_i = \vert \lambda_i - \lambda_i^{\mathrm{NN}} \vert$ \cite{akemann_universal_2019}. An unfolding procedure described in \cite{Prasad_2022} is applied on these $s_i$ to calculate scaled spacings $s_i^{\prime}$. The NNSD is then calculated using the scaled spacings $s_i^\prime$. We provide details of this unfolding procedure in appendix~\ref{app:B}, and, for the sake of brevity, drop the prime notation for the scaled spacings henceforth. Since it is conjectured that the NNSD statistics transitions from Poissonian to a GOE (closed) or GinUE (open) in a system changing from regular to chaotic dynamics, the following quantity can be defined to track this change in the distribution \cite{Prasad_2022},
\begin{align}
    \eta = \frac{\int_0^\infty ds [P(s)-P_{\text{Poi}}(s)]^2}{\int_0^\infty ds [P_{\text{Poi}}(s)-P_{\text{RMT}(s)}]^2} \label{eq:etadefn}
\end{align}
where $P_{\text{RMT}}(s)$ is NNSD of the random matrix theories in question (GOE or GinUE) and $P_{\text{Poi}}(s)$ is that of Poisson distribution (1D in the case of closed model and 2D in case of open model).

The second spectral indicator of chaos we consider is the k$^{\textrm{th}}$-order level spacing ratio, which, like the NNSD, uses the spacing between energy levels to infer chaotic properties in quantum systems. It is defined as \cite{Atas_2013,Srivastava_2018, Tekur_2018, Wang_2022, Tekur_2024} 
\begin{align}
r^i_k=\min\left(\frac{E_{i+2k}-E_{i+k}}{E_{i+k}-E_i},\frac{E_{i+k}-E_i}{E_{i+2k}-E_{i+k}}\right), \,\, r_k^i\in [0,1],
\label{eq:kthlevspacing}
\end{align}
for real spectra $E_i \in \mathbb{R}$. Evidently, in contrast to NNSD, the level spacing ratio can be directly defined using the bare energy eigenvalues without unfolding.
The average k$^{\textrm{th}}$-order level spacing ratio, denoted as $\avg{r_{k}}$, can then be calculated from Eq.~\eqref{eq:kthlevspacing}. For integrable systems, the nearest neighbor level spacing ratio $\expval{ r_1 }$ takes the value $\expval{ r_1 }_{\mr{Poi}}=2\ln2-1\approx0.386$, and for chaotic systems it takes the value $\expval{ r_1 }_{\mr{GOE}}=4-2\sqrt{3}\approx0.536$ \cite{Wang_2022}. Going beyond the nearest neighbor level spacing ratio, to probe the level repulsion due to the long-range correlations in the spectrum, k$^{\textrm{th}}$-order (with $k>1$) level spacing ratio can be used \cite{Srivastava_2018, Tekur_2018, Tekur_2024}. While the nearest neighbor level spacing ratio has been analyzed previously for the Dicke model \cite{Wang_2022}, we will compute the average k$^{\textrm{th}}$-order level spacing ratio and compare the same with that of the GOE and Poisson distributions. This will provide us with an indicator of chaos that is sensitive to long-ranged level repulsion that can be compared to the spectral form factor that we define in the next sub-section. 

In line with the real spectrum, a complex level spacing ratio can also be defined for \cite{sa_complex_2020,Prasad_2022} from the eigenvalues of the Liouvillian as
\begin{align}
    z_i = r_i e^{i\theta_i} = \frac{\lambda_i^{\mr{NN}}-\lambda_i}{\lambda_i^{\mr{NNN}}-\lambda_i},
\end{align}
with $\lambda_i^\mr{NN}$ and $\lambda_i^\mr{NNN}$ denoting the nearest and next-nearest neighbor eigenvalues of $\lambda_i$. While it has been shown in \cite{Prasad_2022} that the two average $\avg{r}$ and $-\avg{\cos \theta}$ undergo a cross-over from the expected values for 2-D Poissonian distribution to GinUE, we will compute these indicators for different strengths of the cavity damping $\gamma$ to address the question of the concurrence of this transition and the dissipative phase transition at $g = g_{c\gamma}$ given in Eq.~\eqref{eq:gc_cav_damp}.

\subsection{Dynamical Indicators of Chaos} \label{sec:SFF}
\begin{figure*}
    \centering
    \includegraphics[width=0.95\linewidth]{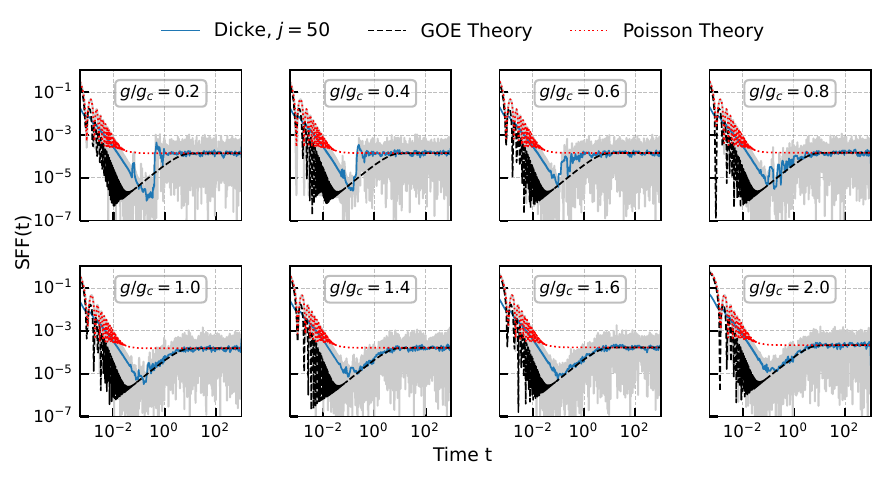}
    \caption{Spectral form factor (SFF) for the closed Dicke model (solid blue line) in the 
    normal/regular ($g/g_c<1$, top panel) and superradiant/chaotic ($g/g_c>1$, bottom panel) regimes. Light grey lines depict the SFF without time averaging, and the black dashed (red dotted) line represents the SFF of the GOE (Poissonian). For all the plots, spin-size $j=50$ and photon cutoff $M=400$ are chosen.}
    \label{fig:fig2}
\end{figure*}

The first dynamical indicator of chaos for closed quantum systems we consider is the spectral form factor (SFF), which has emerged as a powerful tool for probing quantum chaos, especially through the analysis of long-range correlations in the energy spectrum. A hallmark of chaotic systems is the characteristic \textit{dip-ramp-plateau} structure in the SFF, also known as the \textit{correlation hole} \cite{Cornelius_2022, Matsoukas-Roubeas_2023}. The ramp in the SFF, associated with quantum signatures of chaos, arises due to long-range level repulsion between eigenvalues \cite{Cotler_2017}.

First introduced in \cite{Leviandier_1986}, the SFF is defined as the square of the Fourier transform of the two-point correlation function of the spectral density and can be expressed as \cite{Winer_2022}:
\begin{align} 
    \mr{SFF}(t) = \frac{1}{\mathcal{N}^2}\expval{|{\tr e^{i\Hop t} }|^2} = \frac{1}{\mathcal{N}^2} \expval{\left|\sum_{i}e^{iE_it}\right|^2}.
    \label{eq:SFF_defn_1}
\end{align}
Here $\mathcal{N}$ is the total number of eigenvalues or the dimension of the Hilbert space of the Hamiltonian $H$ and the normalization ensures that $\mr{SFF}(t) = 1$ at $t = 0$. Note that, though the SFF as defined above uses the bare energy eigenvalues $E_i$ of the Hamiltonian, in all our calculations of the SFF, we use the unfolded energy eigenvalues $(\{\mathcal{E}_i\})$ in order to extract universal features that are not obscured by local fluctuations in the energy density \cite{Lozej_2023,Nivedita_2020, Buijsman_2020, Hopjan_2023}. Moreover, for quantum systems like the Dicke model, no intrinsic random parameter exists for ensemble averaging. Since the SFF is not a self-averaging quantity, it exhibits strong oscillations over time. To mitigate this, in our calculations, we employ a rectangular kernel for moving time averaging:
\begin{figure*}
    \centering
    \includegraphics[width=0.95\linewidth]{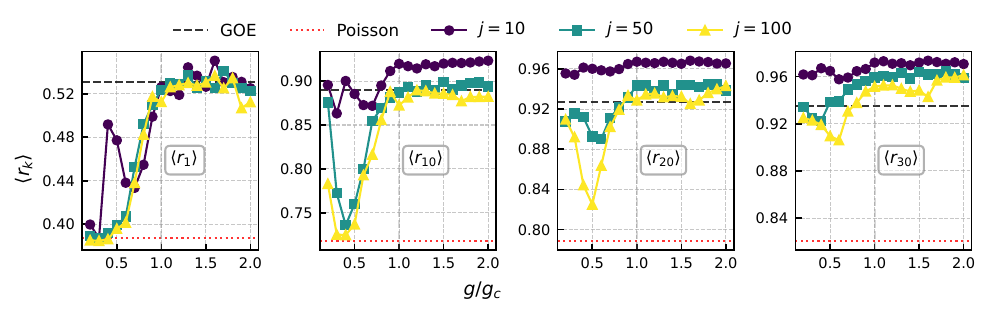}
    \caption{Average k$^{\textrm{th}}$-order level spacing ratio $\avg{r_k}$ (for $k=1, 10, 20, 30$ left to right) as a function of coupling $g$ and varying values of spin size $j$  for the closed Dicke model. The black dashed (red dotted) line represents the level spacing ratio values for the GOE (Poissonian). Photon cutoff is chosen as  $M=400$ in the Dicke model calculation.}
\label{fig:fig3}   
\end{figure*}
\begin{figure}
    \includegraphics[width=0.95\linewidth]{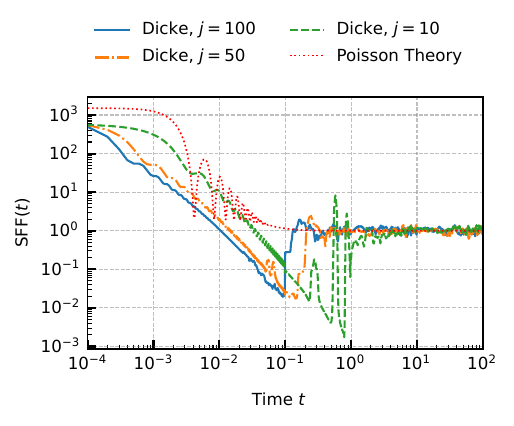}
    \caption{SFF of the closed Dicke model in the normal/regular regime with $g/g_c = 0.4$ for various values of the spin size $j$. The photon number cutoff is chosen as $M=400$ and SFFs for different $j$ are scaled to have the same asymptotic values. The dotted red line is the SFF for the Poissonian model.}
    \label{fig:fig4}   
\end{figure}
\begin{figure*}
    \centering{
    \subfloat{\begin{overpic}[abs,width=0.475\linewidth]{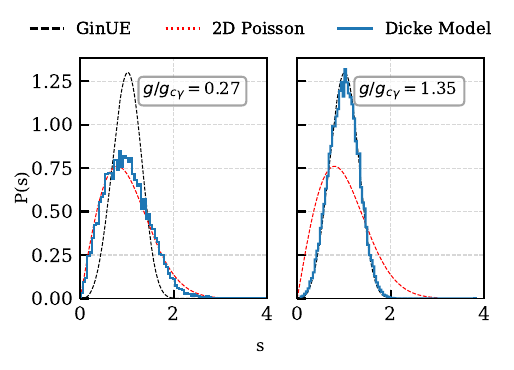}
    \put(110,70){\textcolor{black}{\textbf{(a)}}} \label{fig:fig5a}   
     \end{overpic}
    }
    \subfloat{\begin{overpic}[abs,width=0.475\linewidth]{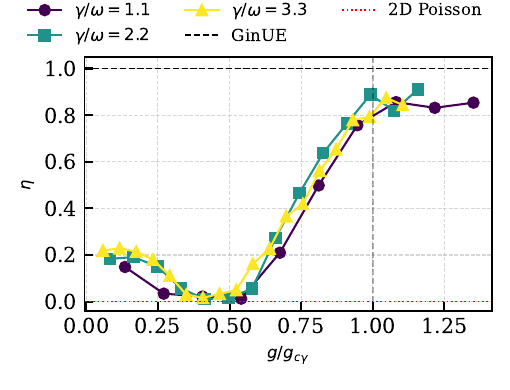}
    \put(70,70){\textcolor{black}{\textbf{(b)}}} \label{fig:fig5b}  
    \end{overpic} 
    }}
    \caption{ (a) NNSD of the complex spectrum of the Liouvillian for the open Dicke model in the normal region with $g/g_{c \gamma} = 0.27$ and in the superradiant region with $g/g_{c \gamma} = 1.35$. Other parameters are $\gamma = 1.1 \omega$ and $j=5$ and photon number cutoff $M=40$. The dashed black (dotted red) line represents the NNSD for the GinUE (2D Poissonian) RMT models. (b) Normalized distance of the open Dicke model's NNSD from the Poissonian NNSD as a function of scaled coupling $g/g_{c\gamma}$ for different values of the cavity damping $\gamma$. Other parameters are the same as in (a).}
    \label{fig:fig5}   
\end{figure*}

\begin{figure}
    \centering
    \begin{overpic}[width=0.95\linewidth]{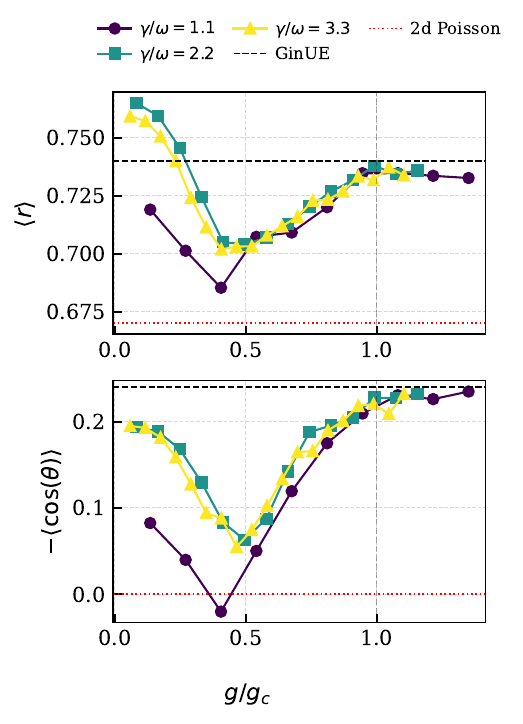}
    \put(40,80){\textcolor{black}{\textbf{(a)}}}
    \put(35,40){\textcolor{black}{\textbf{(b)}}}
    \end{overpic}
    \caption{(a) Radial ($\avg{r}$) and (b) angular ($\avg{\cos(\theta)}$) components of the average complex level spacing ratio of the  Liouvillian spectrum of the open Dicke model as a function of the scaled coupling strength $g/g_{c\gamma}$ and different values of the cavity damping $\gamma$. Spin size is given by $j=5$ and photon number cutoff is $M=40$.}
    \label{fig:fig6}
\end{figure}
\begin{align} 
    \mr{SFF}(t) = \frac{\int_0^{\infty} d\tau \, SFF(\tau) \, \Pi(\tau/\text{win})}{\int_0^{\infty} d\tau \, \Pi(\tau/\text{win})}
    \label{eq:rectangular_smoothing} 
\end{align}
where $\Pi = 1$ if $|\tau/\text{win}| < 1/2$ and $0$ otherwise. The parameter ``win" determines the averaging window size. 

Finally, a direct extension of the SFF \pvedit{for open system} can also be defined as a dynamical indicator of chaos based on the complex eigenvalues of the Liouvillian superoperator mentioned in Eq.~\eqref{eq:master-eq-Dicke}. This measure, referred to as the dissipative spectral form factor (DSFF), is defined as \cite{Li_2021,Prasad_2022,Li_2024}
\begin{align}
    \mr{DSFF}(t,s)=\frac{\expval{ \left|\sum_ne^{i(x_nt+y_ns)}\right|^2}}{\mathcal{N}^2}
    \label{eq:dsff_def1},
\end{align}
with $x_n$ and $y_n$ denote the real and imaginary parts of the complex eigenvalues $\lambda_n = x_n + i y_n$ of the Liouvillian superoperator. The conjugate variables $(t,s)$ can also be organized into a complex variable $p = t + i s = \tau e^{i\varphi}$ leading to a slight modification of Eq.~\eqref{eq:dsff_def1} as
\begin{align}
     \mathrm{DSFF}(\tau,\varphi)=\frac{\expval{\left|\sum_n e^{i(x_n\cos\varphi+y_n\sin\varphi)\tau}\right|^2}}{\mathcal{N}^2}
    \label{eq:dsff_def2}.
\end{align}
In a more recent work by \cite{Li_2024}, it was pointed that the DSFF calculation for dissipative quantum chaotic systems are rather sensitive to the nature of the unfolding and filtering process applied to the eigenvalues of the Liouvillian. They have developed an unfolding and filtering procedure based on applying a suitable transformation of the Liouvillian eigenvalues that leads to the expected \emph{quadratic} ramp for the DSFF agreeing with the GinUE for multiple examples of dissipative quantum chaotic systems. We use this procedure in our calculations and provide further details in appendix~\ref{app:B}. \pvedit{In appendix \ref{app:A}, we also discuss an additional dynamical indicator of chaos, namely the DSPF, for the open Dicke model.}

\section{Results} \label{sec:results}

\subsection{Closed Dicke Model}
We begin the discussion of our results by first focusing on the closed Dicke model. Note that in all the results of this section, we work at resonance $\omega = \omega_0$. In the seminal work of \cite{Emary_2003}, in agreement with the BGS conjecture \cite{Bohigas_1984}, it was shown that there is a regular-to-chaotic transition as the coupling strength is tuned across $g_c$ from the normal-to-superradiant phase of the Dicke model. For the sake of coherence and completeness, we have reproduced the main result of \cite{Emary_2003} in Fig.~\ref{pic1a} where we plot the NNSD for the Dicke model which clearly shows the transition from a Poisson distribution $P_{\text{Poi}}(s)=e^{-s}$ to a Wigner-Dyson distribution $P_{\text{GOE}}(s)=(\pi s/2)e^{-\pi s^2/4}$, characteristic of the GOE, as the coupling is tuned across $g = g_c$. A quantitative way to evidence this transition is to plot the parameter $\eta$ introduced in Eq.~\eqref{eq:etadefn} to characterize the normalized distance of the obtained NNSD from that of the GOE. In Fig.~\ref{pic1b}, we plot $\eta$ for the Dicke model, and again, we see a clear transition from Poissonian to GOE behavior as the coupling is tuned across $g_c$. While the denominator in the calculation of $\eta$ can be evaluated from analytical expressions for the probability distributions, the numerator is evaluated by approximating the integral as a sum using the numerically determined (discretely sampled) $P(s)$ of the Dicke model. Note that in this calculation and the rest of this sub-section, we will restrict ourselves to the even parity sector of the Dicke model \emph{i.e.} we will only consider eigenstates with eigenvalue $+1$ of the operator $\hat{\Pi}$ to reduce the size of the matrix for diagonalization. Moreover, any numerical calculation will have to be done with a finite cut-off $M$ for the number of photons in the cavity mode. To minimize the errors due to this truncation of Fock space, we calculate the eigenvalues for five values of $M= 240, 280, 320, 360, 400$ and ensure convergence of the eigenvalues. Furthermore, in all our results (for the SFF), after selecting these converged eigenvalues, we choose only the middle 60\% of the eigenvalues for the calculations to reduce the fluctuations in the density of states at the edge of the spectrum.

Thus, we see that the nearest neighbor level statistics that capture short-range correlations in the spectrum provide a clear indicator of the regular-to-chaotic transition for the Dicke model. Given this, a related question is to understand how the spectral form factor (SFF) for the Dicke model behaves as the coupling is tuned. The behavior of the SFF (or the survival probability) has been studied in detail in the superradiant regime $g>g_c$ \cite{Villaseñor_2020, Lerma-Hernandez_2019}. In line with the expectation for a chaotic model \cite{Cotler_2017}, the SFF exhibits the characteristic \textit{dip-ramp-plateau} structure in this regime. Interestingly, the behavior of the SFF in the normal or regular regime ($g<g_c$) and as it is tuned across the Dicke phase transition has not been studied in detail. We do this precisely, and the results are displayed in Fig.~\ref{fig:fig2} where the SFF for the Dicke model for various coupling values $g$, with spin-size $j = 50$ and cavity photon number cut-off $M = 400$, are shown. The grey curves in Fig.~\ref{fig:fig2} represent the SFF without ensemble or time averaging, while the blue curves show the SFF after applying a moving time average using the rectangular kernel (Eq.~\ref{eq:rectangular_smoothing}). The black dashed (red dotted) curve represents the SFF for a GOE (Poissonian ensemble) with the same matrix size as the Dicke Hamiltonian (projected to the even parity subspace) calculated using analytical expressions presented in \cite{Ray_2024}. \textit{A key finding of our paper is that even in the regular phase represented in the top panel of Fig.~\ref{fig:fig2}, a correlation hole like structure with a rather sharp dip and jump to the plateau at a finite Heisenberg time $\tau_\mr{Hei}$ appears}. Note that this behavior is rather different from the one expected for the Poisson RMT (red dashed line in Fig.~\ref{fig:fig2}). As we discuss in detail below, this behavior tallies with the persistence of long-range correlations (manifested in the behavior of the average k$^{\mr{th}}$ level spacing ratio) in the spectrum of the Dicke model for finite values ($j=N/2$) of the collective atomic spin. While this makes the behavior of the SFF in the regular and chaotic phase somewhat similar, we note that an important distinguishing feature in the chaotic region with $g>g_c$ is the emergence of a smooth \emph{dip-ramp-plateau} structure with a universal slope in close agreement with the  SFF of GOE as we see in the bottom panel of Fig.~\ref {fig:fig2}.

In order to understand the behavior of the SFF in the regular phase further, we first note that in contrast to the NNSD, which captures short-range level repulsion, it is known that the ramp in the SFF arises from long-range correlations in the spectrum \cite{Shir_2024}. These long-range correlations in the spectrum can also be quantified using the average k$^{\mr{th}}$ level spacing ratio, $\avg{r_k}$ introduced in Eq.~\eqref{eq:kthlevspacing}. In Fig.~\ref{fig:fig3} we plot $\avg{r_k}$ as a function of $g$ for $k=1,10,20,30$ and various values of the spin size $j$. As evident from the figures, the average nearest neighbor spacing shows a clear crossover from Poisson RMT prediction to GOE prediction as $g$ is tuned through $g_c$ for even small values of $j$. This is in agreement with the behavior of the NNSD in Fig.~\ref{pic1a}. In contrast, the higher $k$ level spacing ratios require a very large value of $j$ to show the same behavior. In fact, for the largest value of $j$ \emph{i.e.} $j=100$ we have used in our calculations, while $\avg{r_{10}}$ shows a cross-over from Poisson to GOE but $(\avg{r_{20}},\avg{r_{30}})$ do not. This indicates that we need to approach the thermodynamic limit $j \rightarrow \infty$ even more closely to lose entirely the long-range correlations present in the regular of the Dicke model. Note that the average k$^{\mr{th}}$ level spacing ratio (Eq.~\eqref{eq:kthlevspacing}) can also be defined with unfolded energies. We find that such a redefinition leads to the same qualitative behavior as in Fig.~\ref{fig:fig3}. To bring out this persistence of long-range correlations for even very large values of $j$, in Fig.~\ref{fig:fig4}, we further plot the SFF for the Dicke model in the normal phase with $g/g_c=0.4$ for different values of $j$ (with appropriate rescaling to compare the curves). As we can see from Fig.~\ref{fig:fig4}, the time $t_\mr{Hei}$ at which the SFF tends to its steady state value tends to become smaller as $j$ is increased. \textit{Thus, we may conclude that for the Poissonian random matrix behavior to emerge in the regular region of the Dicke model, fully requires one to approach the thermodynamic limit, $j\rightarrow \infty,$ very closely}. Note that the k$^\mr{th}$ average spacing ratios for the Poisson ensemble were calculated by randomly drawing eigenvalues using its NNSD $P(s)=e^{-s}$. 

Thus, our results indicate that in the normal region of the Dicke model the NNSD agrees with Poissonian RMT, as known previously, there are persistent long-range spectral correlations that do not agree with Poissonian RMT predictions. The physical origin of such long-range spectral correlations in a regime which can be considered regular as per the BGS conjecture is an interesting question whose detailed analysis is beyond the scope of this paper. Nevertheless, we make some remarks towards answering the same. While the results presented above are for the resonant case ($\omega=\omega_0$), we have checked that the results remain qualitatively the same for the off-resonant case as well ($\omega \neq \omega_0$). Thus, our observations are not dependent on fine tuning the parameters of the Dicke model. Moreover, the Dicke model is not Yang-Baxter integrable or Braak criterion integrable for $N \geq 2$ spins for general values of system parameters (normal or superradiant regime) \cite{braak_solution_2013,Batchelor2015,Larson_2017}. To understand if the lack of integrability in this sense may have some implications for the long-range correlations, we can consider the Tavis-Cummings (TC) model, which is the integrable cousin of the Dicke model that also hosts a normal to superradiant transition. While we present the details in Appendix \ref{app:C}, we note here that even for the TC model both the SFF (Fig.~\ref{fig:figAppenC1}) and the average level spacing ratio $\avg{r_k}$ (Fig.~\ref{fig:figAppenC2}) show behaviour not in agreement with the Poissonian RMT in the normal region ($g<g_c$). Thus, the lack of Yang-Baxter integrability is not responsible for long-range correlations. We will revisit this issue in the conclusion of the paper and present some additional suggestions. In the next section, we consider the open Dicke model.

\subsection{Open Dicke Model}

We begin our examination of chaos in the open Dicke model by briefly recapitulating and reproducing (for the sake of completeness) some known results obtained from examining the spectrum of the Liouvillian. For all the results presented in this section, following the same procedure as described in \cite{Prasad_2022} (see appendix~\ref{app:B} for details), we select well converged (with respect to photon number cutoff $M$) complex eigenvalues of the Liouvillian of the open Dicke model. In agreement with them, as shown in Fig.~\ref{fig:fig5a} we find that the NNSD (scaled and smoothed in the manner described in the previous section following \cite{Prasad_2022} and described in appendix~\ref{app:B}) has a 2-D Poissonian behavior for $g<g_{c\gamma}$ and GinUE behavior (with cubic repulsion near $s\sim 0$) for $g>g_{c\gamma}$. 

An interesting open question here is whether the point of transition from 2-D Poissonian-to-GinUE behavior for the open Dicke model coincides with the dissipative normal-to-superradiant quantum phase transition (QPT) that occurs at $g=g_{c\gamma}$. An unequivocal answer to this question would require the calculation of the eigenvalues of the Liouvillian of very high dimension to approach the thermodynamic limit ($N\rightarrow \infty$). For even $N$ as small as $10$, a significantly large photon number of $M=40$ is required to obtain clear statistics for the eigenvalue spacing distribution \cite{Prasad_2022}. Thus, going to even larger $N$ is prohibitively expensive numerically. Nonetheless, we provide an indirect method to understand the connection between the dissipative QPT and the transition from 2-D Poissonian to GinUE statistics. In Fig.~\ref{fig:fig5b} we plot the measure, $\eta$, defined in \cite{Prasad_2022}, measuring the normalized distance between the obtained NNSD distribution for the open Dicke model and the 2D-Poissonian (2D-P) distribution for different values of the cavity decay rate $\gamma$. The fact that the resulting $\eta$ for different values of $\gamma$ collapse onto the same form when plotted as a function of $g/g_{c\gamma}$ (recall definition of $g_{c\gamma}$ from Eq.~\eqref{eq:gc_cav_damp}), indicates that there is a close correlation between the dissipative QPT and the 2D-P to GinUE transition. This is also confirmed by the behavior of the average nearest neighbor complex spacing ratio $\avg{r}$ and $\avg{\cos(\theta)}$ plotted in Fig.~\ref{fig:fig6} for different values of $\gamma$. Interestingly, we find that the crossover from 2D-P to GinUE statistics for the complex spacing ratio is not as monotonic as the behavior shown in Fig.~\ref{pic1b} for $\eta$.

\begin{figure*}
    \centering
    \includegraphics[width=0.95\linewidth]{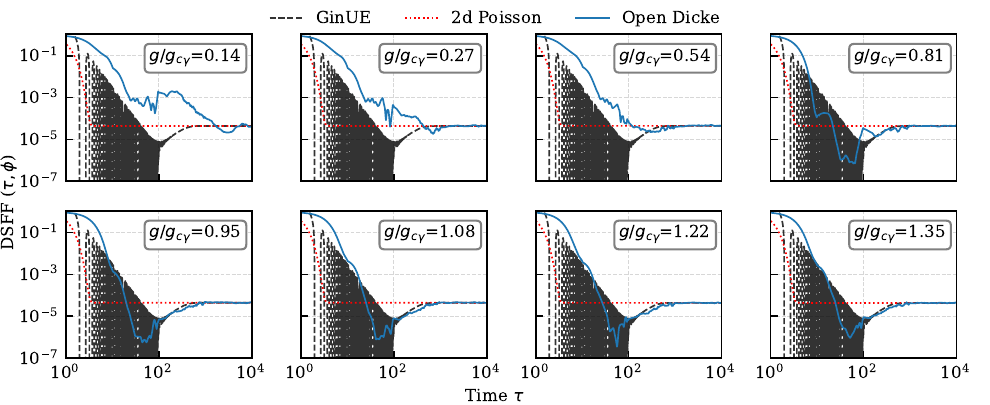}
    \caption{Dissipative spectral form factor (DSFF, blue solid line) for the open Dicke model with cavity damping $\gamma/\omega=1.1$ and $\varphi=3\pi/4$. Spin size is given by $j=5$ and photon number cutoff is taken as $M=40$. The dashed black (dotted red) line represents the DSFF for the GinUE (2D Poissonian) RMT models.}
  \label{fig:fig9}
\end{figure*}
Coming to a dynamical indicator of chaos in the open Dicke model, we consider the DSFF in Eq. (\ref{eq:dsff_def2}), following \cite{Li_2024}, we first unfold the selected complex eigenvalues of the Dicke Liouvillian using the transformation given in Eq.~\eqref{eq:unfolding_transformation} (see appendix~\ref{app:B}) with the parameters $A=-i,\nu=1/3$ and $z_0$ chosen as point of maximum DOS. We choose the branch cut to be $(0,\infty]$. In Fig.~\ref{fig:figAppenB1} of appendix~\ref{app:B}, we have shown the Liouvillian eigenvalues before and after unfolding as well as their smoothed-out densities for the open Dicke model for both $g < g_{c\gamma}$ and $g>g_{c\gamma}$. Using the unfolded eigenvalues, we have calculated the DSFF$(\tau,\varphi)$ with $\varphi$ averaged over a small interval $d \varphi$ around $\varphi=3\pi/4$ and present the result in Fig.~\ref{fig:fig9}. \textit{There, we can clearly see that for $g>g_{c\gamma}$, in the superradiant regime of the open Dicke model, the DSFF has a dip-ramp-plateau behavior with the slope in good agreement with the GinUE RMT model of the same size}. In contrast, for $g<g_{c\gamma}$, while we do not find a dip-ramp structure, neither is the DSFF in perfect agreement with the predictions of a 2D Poissonian model. Both the theoretical plots for GinUE DSFF (dashed black) and 2D Poissonian (dotted red) are plotted using Eqs.~(4) and (5) in \cite{Li_2021} for leading order contributions in $\mathcal{N}$. Note that for the systems considered in \cite{Li_2024}, an extra filtering procedure on top of the unfolded spectrum was required to obtain a uniform DOS and, hence, the desired ramp. In contrast, we find that the unfolded eigenvalues themselves for the different parameter regimes of the open Dicke model considered in Fig.~\ref{fig:fig9}, without any extra filtering, produce a dip-ramp-plateau structure. Thus, in contrast to the behavior of the SFF in the closed case, the DSFF displays distinct behavior in the normal and superradiant phases of the Dicke model and can be considered as a faithful dynamical indicator of the transition of the Liouvillian eigenvalue statistics from 2D-Poissonian to GinUE behavior.

\section{Conclusion} \label{sec:conclusion}

We have demonstrated that the nearest-neighbor spacing distribution (NNSD), the level spacing ratio, and the spectral form factor (SFF) accurately capture the chaotic superradiant region of the closed Dicke model. In agreement with previous results, the structure of the NNSD and the SFF in the chaotic phase aligns well with the behavior predicted by the Gaussian Orthogonal Ensemble (GOE), affirming the presence of quantum chaos in this region. However, our findings challenge a common interpretation in the literature: the mere presence of a correlation hole or a dip-ramp-plateau structure in the SFF does not unconditionally signal chaos. This structure also appears in the regular region of the Dicke model for finite values of $N$, which suggests that it alone is insufficient as a chaos indicator, especially since how closely one has approached the thermodynamic limit can vary between different systems and indicators of chaos. Thus, we conclude that for chaos to be convincingly identified, in addition to the presence of a correlation hole, the structure must closely resemble that of a random matrix ensemble, in accordance with the Bohigas-Giannoni-Schmit (BGS) conjecture.

In the open Dicke model, by examining the NNSD and complex spacing ratio for different values of the cavity damping, we have provided indirect evidence for the concurrence of the dissipative steady-state phase transition and the 2D Poisson-to-GinUE transition as indicated by the Liouvillian eigenvalue statistics. Moreover, we find that a characteristic \emph{dip-ramp-plateau} in agreement with the GinUE RMT prediction is present in the DSFF for the superradiant regime but absent in the normal regime. Thus, we conclude that the DSFF is a sensitive and reliable indicator of of the chaotic phase transition (albeit for transient chaos \cite{Mondal_2025}) in the open Dicke model, affirming its utility in detecting quantum chaos.

We note that the open version of the Dicke model has been implemented in set-ups using ultracold atoms in optical cavities \cite{baumann_dicke_2010,cqedDickequench,DickeRaman} and superconducting qubits in waveguides \cite{mlynek_observation_2014,zanner_coherent_2022,collthermokirchmair}. In the former case \cite{baumann_dicke_2010,cqedDickequench,DickeRaman}, the superradiant phase transition (hence covering the entire parameter regime of interest in our work) and dynamics around it have been experimentally studied. While there are significant challenges such as initial CGS-like state preparation in order to measure SFF or DSFF experimentally, recent work \cite{proposaldetchaos25} indicates that one can indeed consider observables that are closely related to the SFF (which is the survival probability of the hard-to-prepare coherent Gibbs state) such as the survival probability of a simple initial state under quenches. Such variables show features like the correlation hole that is indicative of quantum chaos. Identifying such observables, as well as proposing experimental implementations to measure them in cold atoms systems, where quench dynamics around the Dicke phase transition have been studied \cite{cqedDickequench}, will be an interesting direction of future research. Furthermore, a recent demonstration measuring the SFF in a superconducting quantum processor using the randomized measurement toolbox \cite{SFFQProcessorExpt} also presents a promising approach to experimental realization of the ideas presented here.

As discussed earlier, the physical origin of persistent long-range spectral correlations in the regular regime of the Dicke model can be a subject of future study. While we have ruled out integrability or fine-tuning as possible reasons, a feature common to both the TC and Dicke model where this behavior is observed is the permutation invariance of the spin-cavity interaction. It would be interesting to consider the impact of breaking this symmetry on spectral correlations. Last but not least, performing a further comparative analysis of other dynamical indicators of quantum chaos, beyond the ones considered here, will be interesting. One such measure of interest is the quantum Fisher information (QFI) for the system under consideration in this paper, as it captures an intricate interplay between quantum chaos and multipartite entanglement. In addition, it will also be interesting to generalize the study of QFI in Krylov space as presented in \cite{2024arXiv240711822S,2025arXiv250203434C} for our case. Furthermore, 
given the intricate relationship between time-averaged spread complexity (in Krylov space) and higher-order level spacing ratio discovered in \cite{2025arXiv250414362F}, one can also consider these measures for the Dicke model presented here.

\subsection*{Data Availability}
The codes used to produce the results in this work are available in the following GitHub repository \cite{PrasadGithub2025}.

\begin{acknowledgments}
We thank Adolfo del Campo, Chethan Krishnan, Harsh Sharma, Mahaveer Prasad, Manas Kulkarni, Prithvi Narayan, Sayan Choudhury, and Swathi T S for useful discussions and comments. We acknowledge the use of PARAM ANANTA Supercomputer, commissioned by the National Supercomputing Mission (NSM) for providing computing resources for the HPC
System, which is implemented by C-DAC and supported by the Ministry of Electronics and Information Technology (MeitY) and the Department of Science and Technology (DST), Government of India. 
\begin{figure}
    \centering
    \includegraphics[width=0.95\linewidth]{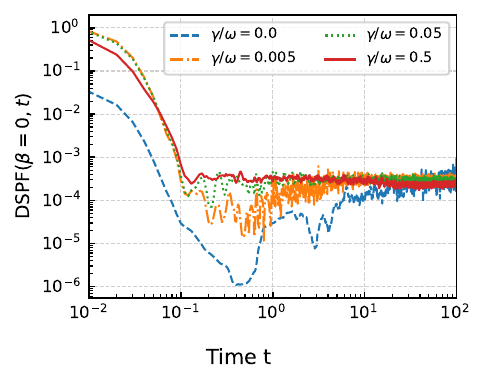}
    \caption{Dissipative survival probability (DSPF) of the open Dicke model for various values of cavity decay $\gamma$ compared to the SFF of the closed Dicke model (blue dashed line). The coupling $g/g_c = 2.0$ is chosen to be in the superradiant regime of the closed model, with spin size $j=20$, photon number cutoff $M=80$, and  $\beta\omega = 0$.}
    \label{fig:figAppenA1} 
\end{figure}
\begin{figure*}
    \centering
 	\begin{overpic}[abs,width=0.9\linewidth]{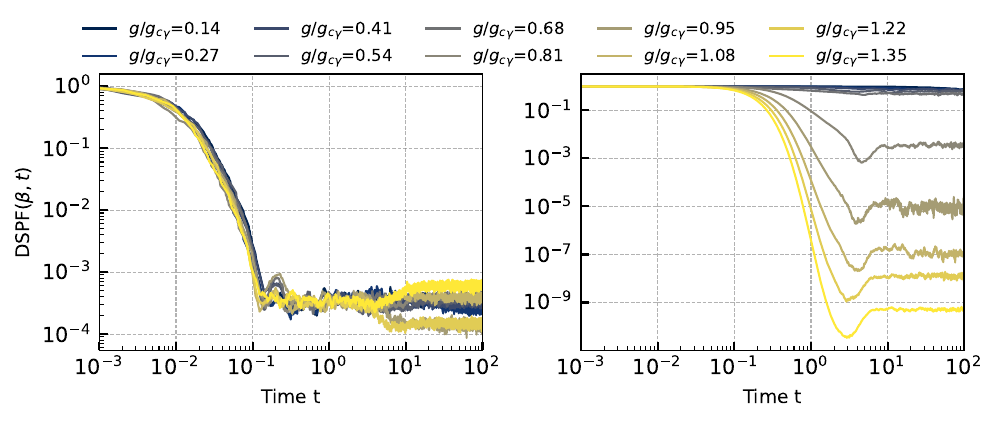}
 		\put(140,120){\textcolor{black}{\textbf{(a)}}}
 		\put(370,120){\textcolor{black}{\textbf{(b)}}}
 	\end{overpic}
    \caption{DSPF of the open Dicke model for various values of coupling $g$ and initial CGS inverse temperature $\beta = 0$ (a) and $\beta = 5/\omega$ (b). The cavity damping rate is chosen as $\gamma/\omega=1.1$, the spin size is $j=20$, and the photon number cutoff is $M=80$. }
    \label{fig:figAppenB2}   
\end{figure*}
PP acknowledges support from a PhD fellowship at IIT Gandhinagar funded by the Ministry of Education, India. AB is supported by the Core Research Grant (CRG/2023/ 001120) by the Department of Science and Technology Science and Anusandhan National Research Foundation (formerly SERB), Government of India. AB also acknowledges the associateship program of the Indian Academy of Sciences (IASc), Bengaluru and would like to thank the Department of Physics of BITS Pilani, Goa Campus, for hospitality during the course of this work, as well as the organizers of the ``Holography, strings and other fun things II" workshop there. AB also acknowledges support from the Indian Institute of Technology Gandhinagar and a generous donor through the Singheswari and Ram Krishna Jha Chair. PV acknowledges support from MATRICS Grant No.~MTR/2023/000900 from Anusandhan National Research Foundation, Government of India. This research was supported in part by the International Centre for Theoretical Sciences (ICTS) by the participation of PP and PV in the program - Quantum Trajectories (code: ICTS/QuTr2025/01).
\end{acknowledgments}

\appendix

\section{DSPF of open Dicke Model}\label{app:A}

\pvedit{Following \cite{Lerma-Hernandez_2019,Villaseñor_2020,Cornelius_2022,Apollonas_2023}, the SFF can also be interpreted as the survival probability, under unitary evolution, of the (infinite temperature $\beta = 0$) Coherent Gibbs State (CGS) $\hrho_\beta = \ketbra{\psi_\beta}$ with }
\begin{align}
\ket{\psi_\beta}
= \sum_n \frac{e^{-\beta E_n/2}}{\sqrt{Z(\beta)}}\ket{n}.\label{eq:CGSState}
\end{align}
\pvedit{This interpretation allows us to define the dissipative survival probability function (DSPF), first suggested in \cite{Tameshtit_1992}, as an analogous indicator of chaos for open quantum systems. The DSPF is defined as \cite{Xu_2021}}
\begin{align}
    \mr{DSPF}(\beta,t) = \expval{\hrho_\beta(t)}{\psi_\beta},
    \label{eq:DSPF}
\end{align}
\pvedit{with $\hrho_\beta(t) = e^{\mathcal{L} t} [\hrho_\beta]$ is the solution of the Lindblad master equation \eqref{eq:master-eq-Dicke} with the initial state given by the CGS. For the open Dicke model, we compte the DSPF numerically by solving the master equation via the Monte Carlo solver \cite{Molmer_1993}  in Python's \textit{QuTip} \cite{qutip2} library. For the results we present here we have obtained the solution by averaging over $100$ trajectories.}
\begin{figure*}
    \centering
    \includegraphics[width=0.9\linewidth]{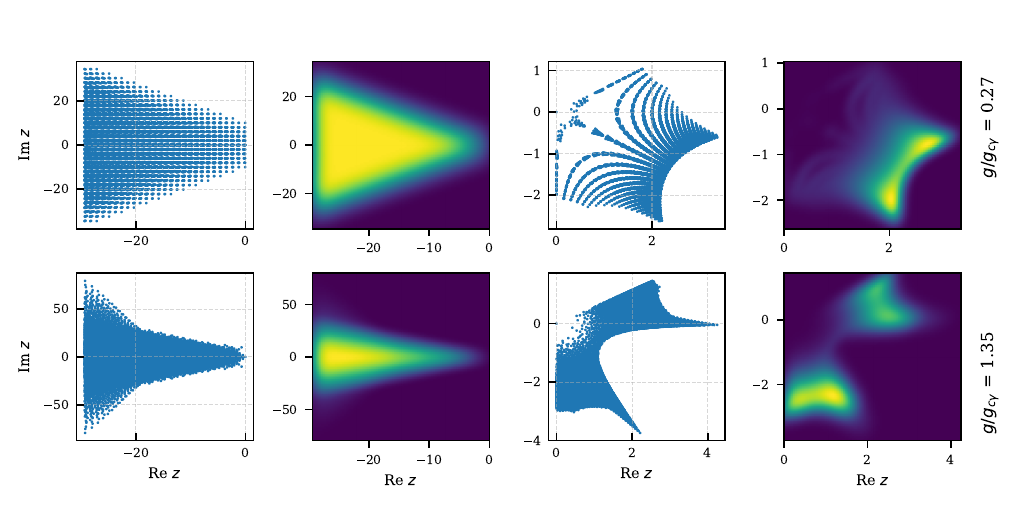}
    \caption{Complex spectrum of the Liouvillian of the open Dicke model for two values of couplings $g$ below (top row) and above (bottom row) critical point $g_{c\gamma}$. The first column shows the scatter plot of the eigenvalues and the second one plots the density heatmap for the same. The third column shows the unfolded eigenvalues as per Eq.~\eqref{eq:unfolding_transformation} and the fourth column plots the corresponding heat map. Spin size is given by $j=5$, photon number cutoff is $M=40$, and cavity damping is given by $\gamma = 1.1 \omega$.}
    \label{fig:figAppenB1}
\end{figure*}
\pvedit{Focusing first on the scenario with the inverse temperature $\beta = 0$ of the initial CGS (see Eq.~\eqref{eq:CGSState}) and $g>g_{c\gamma}$, we see in Fig.~\ref{fig:figAppenA1} that as $\gamma$ is increased, the dip-ramp-plateau structure characteristic of the chaotic regime disappears in agreement with previous results concerning dephasing noise on chaotic models \cite{Xu_2021}. Note that, for $\gamma/\omega=0$, no unfolding of the spectrum is carried out for DSPF \cite{Hopjan_2023}, unlike SFF. Nonetheless, the regular to chaos transition in the close model can impact the DSPF as we see from Fig.~\ref{fig:figAppenB2}, where we plot the DSPF for various values of the coupling strength $g$ and two values of the inverse temperature $\beta$. Here, for the case with $\beta=0$ and the chosen value of $\gamma/\omega = 1.1$, there is no correlation hole for any value of coupling $g$ as expected from Fig.~\ref{fig:figAppenA1}. In contrast, for $\beta \omega=5$ presented in Fig.~\ref{fig:figAppenB2}, we see a correlation hole appears for a large enough coupling strength $g$. Interestingly, we find that the appearance of the correlation hole is not concurrent with the dissipative QPT, which occurs at $g = g_{c\gamma}$. In fact, we find that typically the correlation hole appears in the regime of $g_c<g<g_{c\gamma}$ \emph{i.e.} for coupling values between the closed and open critical coupling strengths. Secondly, the slope of the ramp associated with the correlation hole is not independent of $g$ and does not saturate to the universal slope given by a random matrix model like the GinUE. Furthermore, note that the plateau of the DSPF also does not saturate to a universal value, unlike the SFF, which saturates to $1/\mathcal{N}$. The lack of universality in the slope indicates that the correlation hole of the DSPF for the Dicke model with cavity decay as the dissipative channel is not very sensitive to chaotic properties in the open quantum case.}

\pvedit{Nonetheless, we can understand the behavior qualitatively, since for higher $\beta$, only the ground state and a few excited states have significant weight in the CG state. This, along with the absence of a signature of the dissipative QPT, suggests that the DSPF primarily reflects the transition at $g=g_c$ in the closed model. To support this further, we show in appendix~\ref{app:C} (see Fig.~\ref{fig:figAppenC3}) that for the integrable Tavis-Cummings model that, the DSPF has no correlation hole for any value of $g$. Thus, we find that the DSPF is not particularly sensitive to the dissipative quantum phase transition but still displays a correlation hole structure for initial CGSs with finite inverse temperature $\beta$ in the superradiant region of the closed Dicke model.}

\section{Selecting Eigenvalues and Unfolding the Liouvillian Spectrum} \label{app:B}

\begin{figure}
    \centering
    \includegraphics[width=0.95\linewidth]{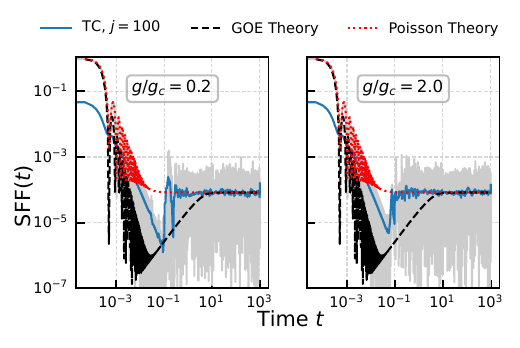}
    \caption{SFF of the closed Tavis-Cummings model in the normal ($g<g_c$) and superradiant ($g<g_c$) regime with spin size $j=100$. Light grey lines depict the SFF without time averaging, and the black dashed (red dotted) line represents the SFF of the GOE (Poissonian) RMT.}
    \label{fig:figAppenC1}
\end{figure}
\begin{figure*}
    \centering
    \includegraphics[width=\linewidth]{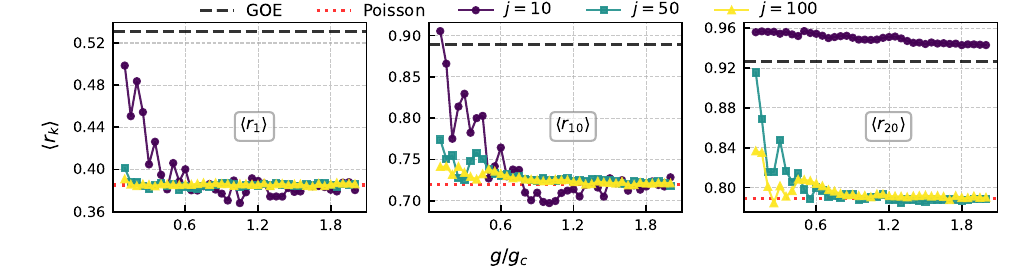}
    \caption{Average k$^{\textrm{th}}$-order level spacing ratio $\avg{r_k}$ (for $k=1, 10, 20$ left to right) as a function of coupling $g$ and varying values of spin size $j$ for the closed Tavis-Cummings model. The black dashed (red dotted) line represents the level spacing ratio values for the GOE (Poissonian). Excitation sectors upto $1000$ were chosen to build the TC model's spectrum.}
    \label{fig:figAppenC2}
\end{figure*}
\begin{figure}
    \centering
    \includegraphics[width=0.95\linewidth]{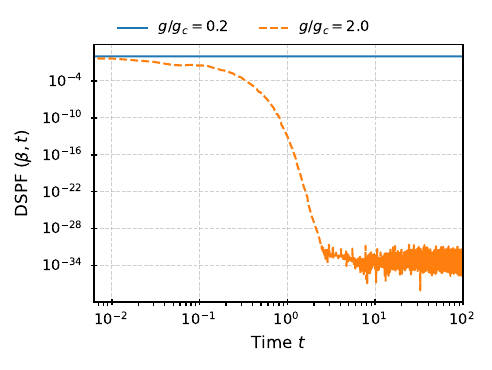}
    \caption{DSPF of the open Tavis-Cummings model for two values of the coupling $g$ above (a) and below (b) the closed model critical point, and initial CGS inverse temperature $\beta = 5/\omega$ (b). The cavity damping rate is chosen as $\gamma/\omega=1.0$, the spin size is $j=20$, and the photon number cutoff is $M=40$.}
    \label{fig:figAppenC3}
\end{figure}
In this appendix, we present the procedure we use to select the eigenvalues of the Liouvillian as well as the two different procedures we use for unfolding them to calculate the NNSD and DSFF based on \cite{Prasad_2022} and \cite{Li_2024}, respectively.

When $\mathcal{N}$ denotes the size of the Hilbert space of the Dicke model, the matrix size of the Liouvillian superoperator is given by $\mathcal{N}^2 \times \mathcal{N}^2$. Thus, it is, in practice, difficult to calculate Liouvillian eigenvalues for large spin size $j$ and photon number cut-off $M$. Following \cite{Prasad_2022}, in our calculations, we take $j=5$ and varying values of $M$, with the largest being $M=40$, and calculate the eigenvalues of the Liouvillian by exact diagonalization of the superoperator. Taking the largest value of $M=40$, in all the results presented in this paper, we select eigenvalues ($z_n$) are selected s.t. Re$(z_n)\in[-\alpha ~\gamma ~M, 0]$. As discussed in \cite{Prasad_2022}, eigenvalues in this range are also typically well converged with respect to changing values of photon number cutoff $M$.

In the first kind of unfolding we use for the calculation of the NNSD \cite{Prasad_2022}, we first determine the nearest neighbor separations $s_i$ for the selected eigenvalues $\lambda_i$ as discussed in the main text. Subsequently, the separations $s_i$ are rescaled to $s_i^\prime = s_i \sqrt{\rho_{\mr{s}}(E_i)}/\bar{s}$ with $\rho_{\mr{s}}(E)$ denoting the smoothened spectral density in the complex plane given by
$\rho_{\rm{avg}}=\frac{1}{2\pi \sigma^2\mathcal{M}}\sum_{i=1}^{\mathcal{M}}
    \exp\left(-\frac{|z-z_i|^2}{2\sigma^2}\right)$
with, $\mathcal{M}$ being the number of complex eigenvalues considered, $\sigma=4.5 ~\tilde{s}$ and $\bar{s}$ is a factor chosen to ensure that the scaled spacings have an average $\sum_{i} s_i^{\prime}/\mathcal{M} = 1$. 

Since the DSFF is rather sensitive to the unfolding procedure used on the Liouvillian eigenvalues, we use an alternative unfolding procedure for its calculation, described in \cite{Li_2024}, which has been shown to lead to the GinUE \emph{dip-ramp-plateau} structure for a variety of scenarios. In particular, we use the following transformation in the complex plane (suggested for non-disordered Hamiltonians in \cite{Li_2024}) 
\begin{align}
    \tilde{z}=g(z)=A(z-z_0)^\nu,
    \label{eq:unfolding_transformation}
\end{align}
to relate the unfolded eigenvalues $\tilde{z}$ and calculated eigenvalues $z$ with $z_0$ being the point in the complex plane where DOS of the spectrum is the maximum. In our calculations we take $A=-i$, $\nu=1/3$. We have plotted the unfolded complex spectra for two different values of couplings $g$, above and below the critical coupling strength $g_{c\gamma}$ along with their respective heatmaps for the DOS in Fig.~\ref{fig:figAppenB1}. Following this transformation, in Eq.~(\ref{eq:dsff_def2}) defining the DSFF, we use $\tilde{x}_n$ and $\tilde{y}_n$ (with $\tilde{z}_n = \tilde{x}_n + i \tilde{y}_n$) instead of $x_n,y_n$ to calculate the DSFF presented in Fig.~\ref{fig:fig9} in the main text.

\section{Tavis-Cummings Model}\label{app:C}

For the purpose of highlighting the connection between the non-integrable nature of the Dicke model \cite{braak_solution_2013,Larson_2017} and the behavior of some of the indicators of chaos that we calculate in this paper, it is also useful to make a comparative study for those indicators in context of the  Tavis-Cummings (TC) model with the Hamiltonian \cite{1968PhRv..170..379T}, 
\begin{align}
    \Hop_\mr{TC} = \omega_0 \Jop_z + \omega \adop \aop + \frac{g}{\sqrt{2j}}(\aop \Jop_+ + \adop \Jop_-) \label{eq:TCH}.
\end{align}
Note that the TC model is closely related to the Dicke model and differs from it due to the rotating wave approximation (RWA) which leads to the conservation of the total excitation number $\hat{Q}=\hat{J_z}+\hat{a}^\dagger \hat{a} + j$ i.e. $\comm{\hat{H}}{\hat{N}}=0$ and hence its integrability. The discrete $\mathbb{Z}_2$ symmetry of the Dicke model turns into $U(1)$ symmetry due to the RWA \cite{Larson_2017}. As a result of this symmetry, we observe line crossings in the spectrum. Thus, we expect regular or non-chaotic behavior for the spectral measures of chaos discussed in Sec.~(\ref{sec:chaos measures}) for any value of $g$ in the TC model. Nonetheless, the closed TC model, like the Dicke model, exhibits a superradiant phase transition with critical coupling strength given by $g_c=\sqrt{\omega\omega_0}$ \cite{Larson_2017}. In Fig.~\ref{fig:figAppenC1}, we plot the SFF for the closed TC model and see that in both the normal and superradiant regimes, the SFF does not agree with the GOE, indicating its non-chaotic nature. Interestingly, a small correlation hole persists even for $g/g_c=0.2$ signalling the presence of long-range correlations in the spectrum. We obtain further evidence for the same from the level spacing ratio average $\expval{r_k}$ as shown in Fig. \ref{fig:figAppenC2} where even for large values of $j$, the behavior of $\expval{r_k}$ for $k=10,20$ do not agree with the Poissonian RMT's prediction for $g<g_c$. Thus, we rule out lack of integrability (or equivalently, presence of counterrotating terms $\hat{a}\hat{J_-}$ and $\hat{a^\dagger}\hat{J_+}$) in the Dicke model to be the reason for existence of long-range correlations. 

In Fig.~\ref{fig:figAppenC3}, we plot the DSPF for the TC model with the initial CGS with finite inverse temperature $\beta \omega = 5$. Again, in contrast to the results for the Dicke model presented in Fig.~\ref{fig:figAppenB2} of the main text, there is no correlation hole for any value of $g$ in this case. Since the initial CGS depends on the closed TC model eigenstates, this contrast, at least in part stems from the integrability of the closed TC model. Moreover, the open TC model also does not have a dissipative QPT and in fact has the vacuum state of the cavity and Dicke state $\ket{j=N/2,m = -N/2}$ with the lowest spin projection as the unique steady state for any value of the coupling $g$ \cite{Larson_2017}. We also note that the ground and first few excited states in the superradiant regime of the closed TC model have very little overlap with the steady state of the open TC model, hence in Fig.~\ref{fig:figAppenC3} the DSPF asymptotically goes to zero for $g<g_c$.

\bibliography{references}

\end{document}